\documentclass[preprint,showpacs,preprintnumbers,nofootinbib,amsmath,amssymb]{revtex4}
\usepackage{graphicx}
\usepackage{dcolumn}
\usepackage{bm}
\begin{document}
\title{Exact and approximate ensemble treatments\\ of thermal pairing
in a multilevel model}
 \author{N. Quang Hung$^{1}$}
 \altaffiliation[On leave of absence from the ]
 {Institute of Physics and Electronics, Hanoi, Vietnam}
  \email{nqhung@riken.jp}
  \author{N. Dinh Dang$^{1, 2}$}
 \email{dang@riken.jp}
\affiliation{1) Heavy-Ion Nuclear Physics Laboratory, RIKEN Nishina
Center for Accelerator-Based Science,
2-1 Hirosawa, Wako City, 351-0198 Saitama, Japan\\
2) Institute for Nuclear Science and Technique, Hanoi, Vietnam}
\date{\today}
\begin{abstract}
    A systematic comparison is conducted for
    pairing properties of finite systems at nonzero
    temperature as predicted by the exact solutions of the pairing problem
    embedded in three principal statistical ensembles, as well as
    the unprojected (FTBCS1+SCQRPA) and Lipkin-Nogami projected
    (FTLN1+SCQRPA) theories that include the quasiparticle
    number fluctuation and coupling to pair vibrations within the
    self-consistent quasiparticle random-phase
    approximation. The
    numerical calculations are performed for the pairing gap, total
    energy, heat capacity, entropy, and microcanonical
    temperature within the doubly-folded
    equidistant multilevel pairing model.
    The FTLN1+SCQRPA predictions agree best with the exact grand-canonical
    results. In general, all approaches
    clearly show that the superfluid-normal phase transition is smoothed
    out in finite systems. A novel formula is suggested for extracting the
    empirical pairing gap in reasonable agreement with the
    exact canonical results.
\end{abstract}

\maketitle

\section{INTRODUCTION}
\label{Intro}
Pairing correlations are a fundamental feature
responsible for the superconducting
(superfluid) properties in many-body systems ranging from
very large ones as neutron stars to tiny ones as atomic nuclei.
In macroscopic systems such as superconductors, pairing
correlations  are destroyed as the
temperature $T$ increases and completely vanish at a
value $T_{\rm c}\simeq 0.57 \Delta(0)$, which is the critical
temperature of the phase transition from the superfluid state to the normal
one. Here $\Delta(0)$ is the value of the pairing gap
at $T=$ 0. The recent years witness a renewed interest in pairing
correlations, which is supported by the exact
solutions of the pairing problem in practice, the studies of unstable
nuclei, the BCS to Bose-Einstein condensation crossover, etc.

The superfluid properties and superfluid-normal (SN) phase
transition of infinite systems are accurately described by the
Bardeen-Cooper-Schrieffer theory~\cite{BCS}, where the average
within the grand canonical ensemble (GCE) is used to obtain the
occupation number in the form of Fermi-Dirac distribution for
noninteracting fermions. The GCE consists of identically prepared
systems in thermal equilibrium, each of which shares its energy and
number of particles with an external heat bath. As compared to the
other two principal thermodynamic ensembles, namely the canonical
ensemble (CE) and the microcanonical one (MCE), the GCE is, perhaps,
the most popular in theoretical studies of systems at finite
temperature because it is very convenient in the calculations of
average thermodynamic quantities~\footnote{For instance, the
double-time Green's functions $G(t,t')$, which are defined by using
the GCE~\cite{Green}, depend only on the time difference $(t-t')$,
which greatly simplify the derivations in many statistical physics
applications.}. The CE is also in contact with the heat bath, but
the particle number is the same for all systems. The MCE is an
ensemble of thermally isolated systems sharing the same energy and
particle number. In thermodynamics limit (i.e. when the system's
particle number $N$ and volume $V$ approach infinity, but $N/V$ is
finite), fluctuations of energy and particle number are zero,
therefore three types of ensembles offer the same average values for
thermodynamic quantities. Thermodynamics limit works quite well in
large systems as well where these fluctuations are negligible. The
discrepancies between the predictions by three types of ensembles
arise when thermodynamics is applied to small systems such as atomic
nuclei or nanometer-size clusters. These systems have a fixed and
not very large number of particles, their single-particle energy
spectra are discrete with the level spacing comparable to the
pairing gap. Under this circumstance, the justification of using the
GCE for these systems becomes questionable. A number of theoretical
studies have also shown that, in these tiny systems, thermal
fluctuations become so large that they smooth out the sharp SN
transition~\cite{Moretto,Goodman,Egido,SPA,Zele,MBCS,FTBCS1,AFTBCS}.
As the result, the pairing gap never collapses but decreases
monotonously with increasing $T$. These predictions are in
qualitative agreement with the results obtained by averaging the
pairing energy in the CE built on the eigenvalues obtained by
exactly solving the pairing problem~\cite{Richardson,EP}. As a
matter of fact, even at $T=$ 0, the exact pairing solution in nuclei
shows a sizable pairing energy in the region where the BCS solution
collapses~\cite{EP}. In the literature so far, under the pretext
that a nucleus is a system with a fixed number of particle, the
thermodynamic averages of the exact solutions of the pairing
Hamiltonian are usually carried out within the CE, and the results
are compared with those obtained within different theoretical
approximations at $T\neq$ 0. The latter, such as the BCS,
Hartree-Fock-Bogoliubov (HFB) theories, etc., as a rule, are always
derived within the GCE, where both energy and particle number
fluctuate. On the other hand, the well-known argument that the
nuclear temperature should be extracted from the MCE of thermally
isolated nuclei is also quite often debated and studied in
detail~\cite{Zele}.

These results suggest that a thorough comparison of the predictions
offered by the exact pairing solutions averaged within three principal
thermodynamic ensembles, and those given by the recent microscopic
approaches, which include fluctuations around the thermal pairing
mean field in nuclei, might be timely and useful. This question is not
new, but the answers to it have been so far only partial.
Already in the sixties Kubo~\cite{Kubo} drew attention to the
thermodynamic effects in very small metal particles.
Later, Denton et al.~\cite{Denton} used Kubo's assumption to study the difference
between the predictions offered by the GCE and CE for the heat capacity
and spin susceptibility within a spinless equidistant level model for
electrons. Very recently,
the predictions for thermodynamic quantities such as total energy,
heat capacity, entropy, and microcanonical temperature
within three principal ensembles
were studied and
compared in Ref. \cite{Sumaryada} by using
the exact solutions of
an equidistant multilevel model with constant pairing interaction
parameter. However, no results for the pairing gaps as functions of
temperature are reported.

In the present paper, we carry out a systematic comparison of
predictions for nuclear pairing properties obtained by averaging the
exact solutions in three
principal thermodynamic ensembles as well as those offered by recent
microscopic approaches to thermal pairing.
For the latter, we choose the unprojected and particle-number projected
versions of the FTBCS1+SCQRPA, which we recently
developed in Ref. \cite{FTBCS1}. This approach has been rigorously derived
based on the same variational procedure used for the
derivation of the standard BCS theory, taking into account the
effects due to quasiparticle-number fluctuation (QNF) and coupling
to the self-consistent quasiparticle random-phase approximation
(SCQRPA) in addition to those of the QNF.

The paper is organized as follows. The pairing Hamiltonian, its
diagonalization, ensemble treatments of the exact pairing solutions,
the main results of the FTBCS1 and FTBCS1+SCQRPA theories, as well as
their particle-number projected versions, are
summarized in Sec. II. The numerical results obtained within
the doubly-folded equidistant multilevel model~\cite{Richardson}
are analyzed in Sec. III.
The paper is summarized in the last section, where conclusions are
drawn.
\section{Exact solution of pairing Hamiltonian}
\label{exact}
\subsection{Exact solution at zero temperature}
\label{zeroT}
We consider a system of
$N$ particles with single-particle energies
$\epsilon_{j}$, which are generated by particle creation operators
$a_{jm}^{\dagger}$ on $j$th orbitals with shell degeneracies
$2\Omega_{j}$ ($\Omega_{j}=j+1/2$), and interacting via a
monopole-pairing force with a constant parameter $G$.
This system is described by the well-known pairing Hamiltonian
\begin{equation}
    H=\sum_{jm}\epsilon_{j}\hat{N}_{j}-
    G\sum_{jj'}\hat{P}_{j}^{\dagger}\hat{P}_{j'}~,
    \label{H}
    \end{equation}
where the particle-number operator $\hat{N}_{j}$ and pairing operator
$\hat{P}_{j}$ are given as
\begin{equation}
    \hat{N}_{j}=\sum_{m}a_{jm}^{\dagger}a_{jm}~,\hspace{5mm}
    \hat{P}_{j}^{\dagger}=\sum_{m>0}a_{jm}^{\dagger}a_{j\widetilde{m}}^{\dagger}~,
    \hspace{5mm} \hat{P}_j=(\hat{P}_j^{\dagger})^{\dagger}~,
\label{N&P}
\end{equation}
with the symbol $~~\widetilde{}~~$ denoting the time-reversal operator, namely
$a_{j\widetilde{m}}=(-)^{j-m}a_{j-m}$.
For a two-component system with $Z$ protons and $N$
neutrons, the sums in Eq. (\ref{H}) run also over all $j_{\tau}m_{\tau}$,
$j'_{\tau}m'_{\tau}$, and $G_{\tau}$ with $\tau=(Z,N)$.

The pairing Hamiltonian (\ref{H}) was solved exactly for the first
time in the sixties by Richardson~\cite{Richardson}. By noticing that
the operators $\hat{J}_{j}^{z}\equiv (\hat{N}_{j}-\Omega_{j})/2$,
$\hat{J}^{+}_{j}\equiv \hat{P}_{j}^{\dagger}$, and
$\hat{J}^{-}_{j}\equiv \hat{P}_{j}$ close an SU(2)
algebra of angular momentum, the authors of Ref. \cite{EP} have
reduced the problem of solving the Hamiltonian (\ref{H}) to its exact
diagonalization in the subsets of representations, each of which is
given by a set basis states $|k\rangle\equiv|\{s_{j}\},
\{N_{j}\}\rangle$ characterized by the
partial occupation number $N_{j}\equiv (J_{j}^{z}+\Omega_{j})/2$ and
partial seniority (the number of
unpaired particles) $s_{j}\equiv\Omega_{j}-2J_{j}$
of the $j$th
single-particle orbital. Here $J_{j}(J_{j}+1)$ is the
eigenvalue of the total angular momentum operator $(\hat{J}_{j})^{2}\equiv
\hat{J}^{+}_{j}\hat{J}^{-}_{j}+\hat{J}^{z}_{j}(\hat{J}^{z}_{j}-1)$.
The partial occupation number $N_{j}$ and seniority $s_{j}$
are bound by the constraints of angular momentum algebra
$s_{j}\leq N_{j}\leq 2\Omega_{j}-s_{j}$ with $s_{j}\leq\Omega_{j}$.
In the present paper, we use
this exact diagonalization method to find the eigenvalues ${\cal E}_{s}$
and eigenstates (eigenvectors) $|s\rangle$ of Hamiltonian
(\ref{H}). Each $s$th eigenstate $|s\rangle$ is fragmented over
the basis states $|k\rangle$ according to
$|s\rangle=\sum_{k}C_{k}^{(s)}|k\rangle$ with the total
seniority $s=\sum_{j}s_{j}$, and degenerated by
\begin{equation}
    d_{s}=\prod_{j}\bigg
    [\frac{(2\Omega_{j})!}{s_{j}!(2\Omega-s_{j})!}-
    \frac{(2\Omega_{j})!}{(s_{j}-2)!(2\Omega-s_{j}+2)!}\bigg]~,
    \label{ds}
    \end{equation}
where $(C_{k}^{(s)})^{2}$ determine the weights of the eigenvector
components. The state-dependent exact occupation number
$f_{j}^{(s)}$ on the $j$th single-particle orbital is then
calculated as the average value of partial occupation numbers
$N_{j}^{(k)}$ weighted over the basis states $|k\rangle$ as
\begin{equation}
    f_{j}^{(s)}=\frac{\sum_{k}N_{j}^{(k)}(C_{k}^{(s)})^{2}}
    {\sum_{k}(C_{k}^{(s)})^{2}}=\sum_{k}N_{j}^{(k)}(C_{k}^{(s)})^{2}~,
    \hspace{2mm} {\rm with}\hspace{2mm} \sum_{k}(C_{k}^{(s)})^{2}=1~.
    \label{fjs}
    \end{equation}
\subsection{Exact solution embedded in thermodynamic ensembles}
\label{thermodynamics}
The properties of the nucleus as a system of $N$ interacting fermions
at energy ${\cal E}$ can be extracted
from its level density $\rho({\cal E},N)$~\cite{Bohr}
\begin{equation}
\rho({\cal E}, N)=\sum_{s, n}\delta({\cal E}-{\cal
E}_{s}^{(n)})\delta(N-n)~,
\label{rho}
\end{equation}
where ${\cal E}_{s}^{(n)}$ are the energies of the quantum
states $|s\rangle$ of the $n$-particle system. Applying to the pairing problem,
these energies are the eigenvalues, which are obtained by exactly diagonalizing
the pairing Hamiltonian
(\ref{H}), as has been
discussed in the previous section.
\subsubsection{Grand canonical and canonical ensembles}
Since each system in a GCE is exchanging its energy and particle number
with the heat bath at a given temperature $T=1/\beta$, both of its energy and
particle number are allowed to fluctuate.
Therefore, instead of the level density (\ref{rho}), it is
convenient to use the average value, which is obtained by integrating
(\ref{rho}) over the intervals in
${\cal E}$ and $N$. This Laplace transform of the level density $\rho({\cal
E}, N)$ defines the grand partition function
$Z(\beta,\lambda)$~\cite{Bohr},
\begin{equation}
    {\cal Z}(\beta,\lambda)=\int\hspace{-3mm}
    \int_{0}^{\infty}\rho({\cal E}, N)e^{-\beta({\cal E}-\lambda N)}dNd{\cal
    E}=\sum_{n}e^{\beta\lambda n}Z(\beta, n)~,
    \label{ZGCE}
    \end{equation}
where $Z(\beta, n)$ denotes the partition function of the CE at
temperature $T$ and particle number fixed at $n$, namely
\begin{equation}
    {Z}(\beta, n)=\sum_{s}p_{s}(\beta,n)~,
    \hspace{5mm} p_{s}(\beta,n)=d_{s}^{(n)}e^{-\beta{\cal
    E}_{s}^{(n)}}~.
    \label{ZCE}
    \end{equation}
Within the GCE, the chemical potential $\lambda$ should be chosen as
a function of $T$ so that the average particle number $\langle
N\rangle$ of the system always remains equal to $N$. The summations
over $n$ and $s$ in Eqs. (\ref{ZGCE}) and (\ref{ZCE}) are obtained
by using Eq. (\ref{rho}) taking into account the degeneracy
$d_{s}^{(n)}$ (\ref{ds}) of each $s$th state in the $n$-particle
system, and carrying out the double integration with $\delta$
functions.

The thermodynamic quantities such as total energy ${\cal E}$, heat
capacity $C$ within the GCE and CE are as usual given as
\begin{equation}
    \langle{\cal E}\rangle_{\alpha}=-\frac{{\partial{\ln}{{\bf
    Z}(\beta)_{\alpha}}}}{{\partial\beta}}; \hspace{5mm}
    C^{(\alpha)}=\frac{\partial\langle{\cal E}\rangle_{\alpha}}{\partial
    T}~
    \label{EC}
\end{equation}
where $\alpha$=GC for the GCE and $\alpha$=C for the CE.

The thermodynamic entropy $S^{(\alpha)}_{\rm th}$ is calculated
within the GCE or CE based on the general definition of the change
of entropy (by Clausius):
\begin{equation}
dS=\beta d{\cal E}~.
\label{S}
\end{equation}
By using the differentials of ${\rm ln}{\bf Z}(\beta)_{\alpha}$, one
obtains
\begin{equation}
    S^{(\rm GC)}_{\rm th} = \beta(\langle{\cal E}\rangle_{\rm GC}-\lambda N)
    +{\rm ln}{\cal Z}(\beta,\lambda)~,\hspace{5mm}
    S^{(\rm C)}_{\rm th} = \beta\langle{\cal E}\rangle_{\rm C}
     +{\rm ln}Z(\beta, N)~.
   \label{SGCE&CE}
\end{equation}

The occupation
number $f_{j}$ on the $j$th single-particle orbital is obtained as
the ensemble average of the state-dependent occupation numbers
$f_{j}^{(s)}$, namely
\begin{equation}
    f_{j}^{(\rm
    GC)}=\frac{1}{{\cal Z}(\beta ,\lambda)}\sum_{s,n}f_j^{(s,n)}d_s^n e^{-\beta({\cal E}_s^{(n)}-\lambda n)}
    ~, \hspace{5mm}
    f_{j}^{(\rm C)}=\frac{1}{{Z}(\beta, N)}\sum_{s}f_{j}^{(s,N)}d_s^N e^{-\beta{\cal E}_s^N}~, \label{fj}
\end{equation}
for the GCE and CE, respectively.
\subsubsection{Microcanonical ensemble}
Different from the GCE and CE, there is no heat bath within the MCE.
A MCE consists of thermodynamically isolated systems, each of which
may be in a different microstate (microscopic quantum state)
$|s\rangle$, but has the same total energy ${\cal E}$ and particle
number $N$. Since the energy and particle number of the system are fixed, one should use
the level density (\ref{rho}) to calculate directly the entropy by
Boltzmann's definition, namely
\begin{equation}
S^{(\rm MC)}({\cal E})={\ln}\Omega({\cal E})~,
\label{SME}
\end{equation}
where $\Omega({\cal E})=\rho({\cal E})\Delta{\cal E}$ is the statistical weight, i.e. the number of
eigenstates of Hamiltonian (\ref{H}) within a fixed energy interval
$\Delta{\cal E}$.
The condition of thermal equilibrium then leads to the standard definition of
temperature within the MCE as~\cite{Bohr}
\begin{equation}
\beta=\frac{\partial S^{(\rm MC)}({\cal E})}{\partial{\cal E}}=
\frac{1}{\rho({\cal E})}\frac{\partial\rho({\cal E})}{\partial{\cal
E}}~,
\label{TMCE}
\end{equation}
using which one can build a ``thermometer'' for each
value of the excitation energy ${\cal E}$ of the system.

In practical calculations, to handle the numerical
derivative at the right-hand side of Eq. (\ref{TMCE}),
one needs a continuous energy dependence of $\rho({\cal E})$
in the form of a distribution in Eq. (\ref{rho}). This is realized by
replacing the Dirac-$\delta$ function $\delta(x)$ at the right-hand
side of Eq. (\ref{rho}),
where $x\equiv{\cal E}-{\cal E}_{s}$, with a nascent $\delta$-function
$\delta_{\sigma}(x)$ ($\sigma >$ 0), i.e. a function that becomes
the original $\delta(x)$ in the limit $\sigma\rightarrow$ 0~\footnote
{This replacement is equivalent to the folding procedure for the
average density, discussed in Sec. 2.9.3 of the textbook
\cite{Ring}, taken at zero degree ($M=$ 0) of the
Laguerre polynomial $L_{M}^{1/2}(x)$.}.
Among the popular nascent $\delta$ functions are the Gaussian (or
normal) distribution, Breit-Wigner
distribution~\cite{BW}, and Lorentz (or relativistic Breit-Wigner)
distribution, which are
given as
\[
\delta_{\sigma}(x)_{\rm G}=\frac{1}{\sigma\sqrt{2\pi}}
{\rm e}^{-\frac{x^{2}}{2\sigma^{2}}}~,\hspace{3mm}
\delta_{\sigma}(x)_{\rm BW}=
\frac{1}{\pi}
\frac{\sigma}{x^{2}+\sigma^{2}}~,
\]
\begin{equation}
\delta_{\sigma}({\cal E}-{\cal E}_{s})_{\rm L}=
\frac{1}{\pi}\frac{\sigma{\cal E}^{2}}
{({\cal E}^{2}-{\cal E}_{s}^{2})^{2}+
\sigma^{2}{\cal E}^{2}}~,
\label{distribution}
\end{equation}
respectively. In these distributions, $\sigma$ is a parameter, which
defines the width of the peak centered at ${\cal E}={\cal E}_{s}$. The
full widths at the half maximum (FWHM) $\Gamma$ of these distributions are
$\Gamma= 2\sigma\sqrt{2{\ln} 2}\simeq 2.36\sigma$ for the Gaussian
distribution, and $2\sigma$ for the Breit-Wigner and Lorentz ones.
The disadvantage of using such smoothing
is that the temperature extracted from Eq. (\ref{TMCE}),
of course, depends on the chosen distribution as well as the value of
the parameter $\sigma$. It is worth mentioning that changing $\sigma$
is not equivalent to changing $\Delta E$ for the discrete $\rho({\cal
E})$ used
in calculating the statistical weight $\Omega({\cal E})$ in Eq.
(\ref{SME}), since the wings of any distributions in Eq.
(\ref{distribution}) extend with increasing $\sigma$, whereas for
the discrete spectrum, no more levels might be found
in the low ${\cal E}$ by enlarging $\Delta{\cal E}$.

The generalized form of Boltzmann's entropy (\ref{SME}) is the
definition by von Neumann
\begin{equation}
    S=-{\rm Tr}[\rho{\rm ln}\rho]~,
    \label{Neumann}
    \end{equation}
    which is the quantum mechanical correspondent of the Shannon's
    entropy in classical theory.
    By expressing the level density $\rho({\cal E})$ in Eq. (\ref{rho})
in terms of the local densities of states $F_{k}({\cal E})$~\cite{Zele},
\begin{equation}
    \rho({\cal E})=\sum_{k}F_{k}({\cal E})~, \hspace{5mm}
    F_{k}({\cal E}) = \sum_{s}[C_{k}^{(s)}]^{2}
    \delta({\cal E}-{\cal E}_{s})~,
    \label{Fs}
    \end{equation}
the entropy becomes
\begin{equation}
    S^{(s)}=-\sum_{k}[C_{k}^{(s)}]^{2}\ln[C_{k}^{(s)}]^{2}~.
    \label{Ss}
    \end{equation}
By using Eqs. (\ref{TMCE}) and (\ref{Ss})
one can extract a quantity $T_{s}$ as the ``microcanonical temperature
of each eigenstate $s$''.
\subsection{Determination of pairing gaps from total energies}
\subsubsection{Ensemble-averaged pairing gaps}
Although the exact solution of Hamiltonian (\ref{H}) does not
produce a pairing gap {\it per se}, which is a quantity determined
within the mean field, it is useful to define an ensemble-averaged
pairing gap to be closely compared with the gaps predicted by the
approximations within and beyond the mean field. In the present
paper, we define this ensemble-averaged gap $\Delta_{\alpha}$ from
the pairing energy ${\cal E}_{\rm pair}^{(\alpha)}$ of the system as
follows
\begin{equation}
    \Delta_{\alpha}=\sqrt{-G{\cal E}_{\rm pair}^{(\alpha)}}~,\hspace{5mm}
    {\cal E}_{\rm pair}^{(\alpha)}=\langle{\cal
    E}\rangle_{\alpha}-\langle{\cal
    E}\rangle_{\alpha}^{(0)}~, \hspace{5mm} \langle{\cal
    E}\rangle_{\alpha}^{(0)}\equiv
    2\sum_{j}\Omega_{j}\big[\epsilon_{j}-
    \frac{G}{2}f_{j}^{(\alpha)}\big]f_{j}^{(\alpha)}~,
    \label{gapGC&C}
    \end{equation}
within the GCE ($\alpha={\rm GC}$), CE ($\alpha={\rm C}$), and
MCE ($\alpha={\rm MC}$), where for the latter we put
${\cal E}_{\rm pair}^{(\alpha)}\equiv{\cal E}_{\rm pair}(s)$
with $\langle{\cal
    E}\rangle_{\rm MC}\equiv{\cal
    E}_{s}^{(N)}$.
The term $\langle{\cal E}\rangle_{\alpha}^{(0)}$ denotes the
contribution from the energy
$2\sum_{j}\Omega_{j}\epsilon_{j}f_{j}^{(\alpha)}$ of the
single-particle motion described by the first term at the right-hand
side of Hamiltonian (\ref{H}), and the energy
$-G\sum_{j}\Omega_{j}[f_{j}^{(\alpha)}]^{2}$ of uncorrelated
single-particle configurations caused by the pairing interaction in
Hamiltonian (\ref{H}). Therefore, subtracting the term $\langle{\cal
E}\rangle_{\alpha}^{(0)}$ from the total energy $\langle{\cal
E}\rangle_{\alpha}$ yields the residual that corresponds to the
energy due to pure pairing correlations. By replacing
$f_{j}^{(\alpha)}$ with $v_{j}^{2}$, one recovers from Eq.
(\ref{gapGC&C}) the expression ${\cal E}_{\rm pair}^{(\rm BCS)}
=-\Delta^{2}_{\rm BCS}/G$ of the BCS theory. Given several
definitions of the ensemble-averaged gap existing in the literature,
it is worth mentioning that the definition (\ref{gapGC&C}) is very
similar to that given by Eq. (52) of Ref. \cite{Delft}, whereas,
even within the CE, the gap $\Delta_{\rm C}$ is different from the
canonical gap defined in Refs. \cite{Ross,Frau}, since in the
latter, the term $\langle{\cal E}\rangle_{\rm C}^{(0)}$ is taken at
$G=$ 0. The pairing energy ${\cal E}_{\rm pair}^{(\alpha)}$ in Eq.
(\ref{gapGC&C}) is also different from the simple average value
$-G\sum_{jj'}\langle{\hat{P}^{\dagger}_{j}\hat{P}_{j'}}\rangle_{\alpha}$
of the last term of Hamiltonian (\ref{H}) as the latter still
contains the uncorrelated term
$-G\sum_{j}\Omega_{j}[f_{j}^{(\alpha)}]^{2}$.
\subsubsection{Empirical determination of pairing gap at finite
temperature}
The simplest way to empirically determine the pairing gap 
of a system with $N$ particles in the ground state (at $T=$ 0) is to
use the so-called three-point formula $\Delta^{(3)}(N)$, which is
given by the odd-even mass difference between the ground-state
energies of the $N$-particle system and the neighboring systems with
$N\pm1$ particles~\cite{Bohr}.
A straightforward extension of this formula to $T\neq$ 0 reads
\begin{equation}
    \Delta^{(3)}(\beta,N)
    \simeq \frac{(-1)^{N}}{2}[\langle{\cal E}(N+1)\rangle_{\alpha}-
    2\langle{\cal E}(N)\rangle_{\alpha}+\langle{\cal
    E}(N-1)\rangle_{\alpha}]~,
    \label{Soddeven}
    \end{equation}
which was used, e.g., in Ref. \cite{Kaneko} to extract the thermal
pairing gaps in wolfram and molybdenum isotopes. The four-point
formula represents the arithmetic average of the three-point gaps over in 
the neighboring systems with $N$ and $N-1$ particles, namely
\begin{equation}
    \Delta^{(4)}(\beta,N) = \frac{1}{2}[\Delta^{(3)}(N) +
    \Delta^{(3)}(N-1)]~.
    \label{4point}
    \end{equation}
A drawback of the gaps
$\Delta^{(3)}(\beta,N)$ and $\Delta^{(4)}(\beta,N)$, 
defined in this way, is that they still contain the
admixture with the contribution from uncorrelated single-particle
configurations. The later increases with increasing $T$. Therefore,
Eq. \eqref{Soddeven} and, consequently, Eq. (\ref{4point}) do 
not hold at finite temperature. To remove
this contribution so that the experimentally extracted pairing gap
is comparable with $\Delta_{\alpha}$ in Eq. (\ref{gapGC&C}), we
propose in the present paper an improved odd-even mass difference
formula at $T\neq$ 0 as follows. Using Eq. (\ref{gapGC&C}) to
express the total energy $\langle{\cal E}(N)\rangle_{\alpha}$ of the
system in terms of $\Delta_{\alpha}(N)$ and $\langle{\cal
E}(N)\rangle_{\alpha}^{0}$, we obtain
    \begin{equation}
        \langle{\cal E}(N)\rangle_{\alpha}=
    \langle{\cal E}(N)\rangle_{\alpha}^{(0)}
    -\frac{\widetilde{\Delta}^{2}(\beta,N)}{G}~.
  \label{EN}
  \end{equation}
  where $\langle{\cal E}(N)\rangle_{\alpha}$ is the experimentally
  known total energy of the system with $N$ particles
  at $T\neq$ 0, whereas $\widetilde{\Delta}(\beta,N)$ is the pairing gap
  of this system to be determined.
  Replacing $\langle{\cal E}(N)\rangle_{\alpha}$ in the definition of
  the odd-even mass difference (\ref{Soddeven}) with the right-hand
  side of Eq. (\ref{EN}), we obtain a quadratic equation for the
  three-point $\widetilde{\Delta}^{(3)}(\beta,N)$
    \begin{equation}
    \widetilde\Delta^{(3)}(\beta,N)=
    (-1)^{N}\bigg\{\frac{1}{2}[\langle{\cal E}(N+1)\rangle_{\alpha}+\langle{\cal
    E}(N-1)\rangle_{\alpha}]-
    \langle{\cal
    E}\rangle_{\alpha}^0+\frac{[\widetilde\Delta^{(3)}(\beta,N)]^{2}}{G}\bigg\}~.
    \label{gapodd-even}
    \end{equation}
    The discriminant of this equation is equal to
    $G\sqrt{1-4{S'}/{G}}$, where
        \begin{equation}
	 S'=\frac{1}{2}\big[\langle{\cal E}(N+1)\rangle_{\alpha}+\langle{\cal
	 E}(N-1)\rangle_{\alpha}\big]-\langle{\cal
	 E}(N)\rangle_{\alpha}^{(0)}~.\label{S'}
        \end{equation}
Therefore the condition for Eq. (\ref{gapodd-even}) to have real
solutions is $S'\leq G/4$. Including both cases with even and odd $N$,
the positive solution of Eq. (\ref{gapodd-even}) is always possible
provided $S' <$ 0, which reads 
  \begin{equation}
      \widetilde{\Delta}^{(3)}(\beta,N)=
      \frac{G}{2}\bigg[(-1)^{N}+\sqrt{1-4\frac{S'}{G}}\bigg]~.
      \label{gapN}
  \end{equation}  
    The quantity $S'$ differs from the conventional odd-even mass difference
    shown in the square brackets of (\ref{Soddeven}) by the contribution due 
    to the uncorrelated single-particle
    motion, i.e. the last sum containing $G$ in the definition of
    $\langle{\cal E}\rangle_{\alpha}^{(0)}$ in Eq. (\ref{gapGC&C}).
    The latter is zero only at $G=$ 0, which yields 
    $\widetilde{\Delta}^{(3)}(\beta,N)=$ 0, as  can be seen from Eq.
    (\ref{gapN}) as well. In this case both $S'$ and
    the expression in the square brackets of Eq. (\ref{Soddeven}) vanish as they are just 
    the difference of the Hartree-Fock energies
    $\langle E(N+1)\rangle^{(0)}_{\alpha} + \langle
    E(N-1)\rangle^{(0)}_{\alpha}- 2\langle
    E(N)\rangle^{(0)}_{\alpha}$, which is zero at $G = 0$. Moreover,
    while the odd-even mass difference of Eq. (\ref{Soddeven}) can be 
    positive or negative depending on whether $N$ is even or odd, $S'$
    should be always negative as discussed above.
The gap $\widetilde{\Delta}^{(3)}(\beta,N)$ extracted from Eq. (\ref{gapN}) is, therefore,
consistent with the result of the exact calculation at zero
temperature, where the pairing gap is zero only at $G$ = 0, and increases
with $G$ (See e.g. Fig. 1 - (a) of Ref.
\cite{Sumaryada}). As compared to the simple finite-temperature
extension of the odd-even mass (\ref{Soddeven}), the modified gap
$\widetilde{\Delta}^{(3)}(\beta,N)$ is closer to the ensemble-averaged gap
$\Delta_{\alpha}(N)$ (\ref{gapGC&C}) since it is free from the
contribution of uncorrelated single-particle configurations. In Eq.
(\ref{gapN}), the energies $\langle{\cal E}(N+1)\rangle_{\alpha}$
and $\langle{\cal E}(N-1)\rangle_{\alpha}$ can be extracted from
experiments, whereas the pairing interaction parameter $G$ can be
obtained by fitting the experimental values of $\Delta(T=0,N)$. The
energy $\langle{\cal E}(N)\rangle_{\alpha}^{(0)}$ remains the only
model-dependent quantity being determined in terms of the
single-particle energies $\epsilon_{j}$ and single-particle occupation numbers
$f_{j}^{(\alpha)}$.  As a
matter of fact, since
the pairing gap $\Delta^{(3)}(\beta,N)$ of the $N$-particle system 
at the left-hand side of the
expression
(\ref{Soddeven}) is also present in the total energy $\langle{\cal
E}(N)\rangle_{\alpha}$ of the same system at the right-hand
side of (\ref{Soddeven}), the former is simply extracted from the
latter by using Eqs. (\ref{gapGC&C}) and (\ref{EN}). 
As the result, the modified gap $\widetilde{\Delta}^{(3)}(\beta,N)$
of the system with $N$ particles explicitly 
depends now on $\langle{\cal E}(N)\rangle_{\alpha}^{(0)}$ of the same 
system
rather than on its total energy $\langle{\cal E}(N)\rangle$. The
modified four-point gap $\widetilde{\Delta}^{(4)}(\beta,N)$ is then
obtained from the modified three-point gaps $\widetilde{\Delta}^{(3)}(\beta,N)$
and $\widetilde{\Delta}^{(3)}(\beta,N-1)$ by using the definition
(\ref{4point}).
\section{FTBCS1+SCQRPA and FTLN1+SCQRPA}
The FTBCS1+SCQRPA includes the effects due to
quasiparticle-number fluctuation and coupling to the SCQRPA
vibrations, which are neglected within the standard BCS theory.
The derivation of the FTBCS1+SCQRPA has already been given
and discussed in detail in Ref~\cite{FTBCS1}.
Therefore we give below only the main
results, which are necessary to follow the numerical
calculations in the present paper.

The rigorous derivation of the
FTBCS1+SCQRPA theory follows the standard variational procedure, which is
used to derived the BCS theory. By using the Bogoliubov's canonical
transformation~\cite{Bogo}
\begin{equation}
{\alpha}_{jm}^{\dagger}=u_{j}a_{jm}^{\dagger}-v_{j}a_{j\widetilde{m}}~,
\hspace{5mm}
 {\alpha}_{j\widetilde{m}}=u_{j}
 a_{j\widetilde{m}}+v_{j}a_{jm}^{\dagger}~,
 \label{Bogotransform}
 \end{equation}
the pairing Hamiltonian (\ref{H}) is expressed in terms of
quasiparticle operators, $\alpha_{jm}^{\dagger}$ and $\alpha_{jm}$,
as $H_{\rm q.p.}$, whose explicit form is given in many papers,
e.g., Eqs. (3), (8) -- (14) of Ref. \cite{FTBCS1}. The $u_{j}$ and
$v_{j}$ coefficients of the Bogoliubov's transformation
(\ref{Bogotransform}) are determined by minimizing the GCE average
value of the Hamiltonian ${\cal H}=H_{\rm q.p.}-\lambda\hat{N}$.
This leads to the equation
\begin{equation}
\langle[{\cal H},{\cal A}_{j}^{\dagger}]\rangle_{\rm
GC}=0~,\hspace{5mm} {\rm where}\hspace{2mm}
{\cal A}_{j}^{\dagger}=\frac{1}{\sqrt{\Omega_{j}}}
        \sum_{j=1}^{\Omega_{j}}\alpha_{jm}^{\dagger}
        \alpha_{j\widetilde{m}}^{\dagger}~.
        \label{[HA]}
        \end{equation}
The final result yields the equation
for the level-dependent pairing gap $\Delta_{j}$ for a system with
even number $N$ of particles,
        \begin{equation}
       \Delta_{j}=\Delta+\delta\Delta_{j}~,\hspace{5mm}
       \Delta=G\sum_{j'}\Omega_{j'}(1-2n_{j'})u_{j'}v_{j'}~,
       \hspace{5mm}
       \delta\Delta_{j}=
       2G\frac{\delta{\cal N}_{j}^{2}}{1-2n_{j}}u_{j}v_{j}~,
        \label{gapjFTBCS1}
        \end{equation}
with the QNF defined as
\begin{equation}
\delta{\cal N}_{j}^{2}\equiv n_{j}(1-n_{j})~,
\label{QNF}
\end{equation}
and the equation for the average particle number $N$,
\begin{equation}
N=2\sum_{j}\Omega_{j}[v_{j}^{2}(1-2n_{j})+n_{j}]~,
\label{NFTBCS1}
\end{equation}
The $u_{j}$ and $v_{j}$ coefficients in Eqs. (\ref{gapjFTBCS1}) and
(\ref{NFTBCS1})
are determined as
\begin{equation}
u_{j}^{2}=\frac{1}{2}\bigg[1+\frac{\epsilon_{j}'-Gv_{j}^{2}-\lambda}{E_{j}}\bigg]~,\hspace{5mm}
v_{j}^{2}=\frac{1}{2}\bigg[1-\frac{\epsilon_{j}'-Gv_{j}^{2}-\lambda}{E_{j}}\bigg]~,
    \label{uv}
    \end{equation}
with the renormalized single-particle energies $\epsilon_{j}'$,
        \begin{equation}
        \epsilon_{j}'=\epsilon_{j}+\frac{G}{\sqrt{\Omega_{j}}(1-2n_{j})}
        \sum_{j'}\sqrt{\Omega_{j'}}(u_{j'}^{2}-v_{j}^{2})
        (\langle{\cal A}^{\dagger}_{j}{\cal A}^{\dagger}_{j'\neq j}\rangle+
        \langle{\cal A}^{\dagger}_{j}{\cal A}_{j'}\rangle)~,
        \label{renej}
        \end{equation}
and the quasiparticle energies
\begin{equation}
    E_{j}=\sqrt{(\epsilon_{j}'-Gv_{j}^{2}-\lambda)^{2}+\Delta_{j}^{2}}~.
    \label{Ej}
    \end{equation}
For a system with an odd number of particles, the blocking effect
caused by the unpaired particle should be taken into account in  Eqs. (\ref{gapjFTBCS1}) and
(\ref{NFTBCS1}). In the
present paper, for simplicity, we do not consider systems with odd
particle numbers within the FTBCS1+SCQRPA and FTLN1+SCQRPA. 

The pair correlators $\langle{\cal A}^{\dagger}_{j}{\cal
A}^{\dagger}_{j'\neq j} \rangle$ and $\langle{\cal
A}^{\dagger}_{j}{\cal A}_{j'}\rangle$ in Eq. (\ref{renej}) are
determined by numerically solving a set of coupled equations (47)
and (48) of Ref. \cite{FTBCS1}, which contain the ${\cal
X}_{j}^{\mu}$ and ${\cal Y}_{j}^{\mu}$ amplitudes of the SCQRPA
equations. The details of the derivation of the SCQRPA are given in
Ref. \cite{SCQRPA}. The SCQRPA equations are solved
self-consistently with the gap and number equations
(\ref{gapjFTBCS1}) and (\ref{NFTBCS1}). The quasiparticle occupation
numbers $n_{j}$ are then found by solving a set of equations that
include coupling of quasiparticle density operators
$\alpha_{jm}^{\dagger}\alpha_{jm}$ to the SCQRPA phonon operators.
The result yields the integral equation (69) of Ref. \cite{FTBCS1}
for $n_{j}$. To compare with the level-independent gap such as the
BCS one, the level-weighted gap,
\begin{equation}
\overline{\Delta}=\sum_{j}\Omega_{j}\Delta_{j}/\sum_{j}\Omega_{j}~,
\label{gapFTBCS1}
\end{equation}
is used instead of Eq. (\ref{gapjFTBCS1}). The total energy
$\langle{\cal E}\rangle_{\rm FTBCS1+SCQRPA}$ is calculated by
averaging the quasiparticle representation $H_{\rm q.p.}$ of the
Hamiltonian (\ref{H}) within the GCE, i.e.
\begin{equation}
    \langle{\cal E}\rangle=\langle H_{\rm q.p.}\rangle_{\rm GC}~,
    \label{<E>}
    \end{equation}
and the heat capacity $C$ is then found from Eq. \eqref{EC}. The
thermodynamic entropy is obtained by integrating Eq. \eqref{S}
\begin{equation}
    S_{\rm th}=\int_{0}^{T}\frac{1}{\tau}C d\tau~.
    \label{SthBCS}
\end{equation}
The BCS equations~\cite{BCS,Bogo} are recovered from Eqs.
(\ref{gapjFTBCS1}) and (\ref{NFTBCS1}) by neglecting the QNF
$\delta{\cal N}_{j}^{2}$ [i.e. $\delta\Delta_{j}=0$ in Eq.
(\ref{gapjFTBCS1})] together with the pair correlators $\langle{\cal
A}^{\dagger}_{j}{\cal A}^{\dagger}_{j'\neq j} \rangle$ and
$\langle{\cal A}^{\dagger}_{j}{\cal A}_{j'}\rangle$ in Eq.
(\ref{renej}), and assuming $n_{j}$ to have the form of Fermi-Dirac
distribution for noninteracting quasiparticles, i.e. setting
\begin{equation}
n_{j}=n_{j}^{\rm FD}\equiv\frac{1}{e^{\beta E_{j}}+1}~.
\label{FDn}
\end{equation}

The BCS is known to violate the particle number. This causes a
certain quantal particle-number fluctuation around the average value
determined by Eq. (\ref{NFTBCS1}) even at $T=$ 0. The FTBCS1+SCQRPA
takes into account only the thermal effect in terms of QNF
$\delta{\cal N}_{j}^{2}$, but does not remove the quantal
fluctuation, which is a feature inherent in the BCS wave functions.
To cure this inconsistency, a proper particle-number projection
(PNP) needs to be carried out. The Lipkin-Nogami (LN)
method~\cite{LN} is an approximated PNP before variation, which is
widely used in nuclear study because of its simplicity. This method
has been implemented into the FTBCS1+SCQRPA, and the ensuing
approach is called the FTLN1+SCQRPA (See Sec. II.C.2. of Ref.
\cite{FTBCS1}). However, the LN method can approximately eliminate
only the quantal fluctuations due to particle-number violation 
within the BCS theory. These quantal
fluctuations are different from the thermal particle-number
fluctuations, which always arise from the exchange of particles 
between the systems in the GCE. The LN method is, therefore, not sufficient
to remove the particle-number fluctuations within the GCE. 
To avoid the thermal particle-number fluctuations, the average in the
CE should be used instead. Unfortunately, the methods of equilibrium
statistical physics applied to the nuclear theories such as the
Matsubara-Green's function and/or double-time Green's function
techniques, which are used to derived the BCS and QRPA equations at
finite temperature, are 
all based on the GCE. The complete particle-number projection based on
the applying particle-number projection operator~\cite{PNP} at
finite temperature for these approaches still remains a subject under study 
[See, e.g. Ref. \cite{PMBCS}].

In this paper, the numerical results obtained within both
particle-number unprojected (FTBCS1+SCQRPA) and projected
(FTLN1+SCQRPA) approaches will be compared with those given by
averaging the exact pairing solution with the principal
thermodynamic ensembles.
\section{NUMERICAL RESULTS}
\label{results}
\subsection{Details of numerical calculations}
The schematic model employed for numerical calculations consists of
$N$ particles, which are distributed over $\Omega=N$ doubly-folded
equidistant levels (i.e. with the level degeneracy $2\Omega_{j}=2$).
These levels, whose energies are $\epsilon_{j} =
\epsilon[j-(\Omega+1)/2]$ ($j=$ 1,~\ldots,~$\Omega$), interact via
the pairing force with a constant parameter $G$. The model is
half-filled, namely, in the absence of the pairing interaction, all
the lowest $\Omega/2$ levels (with negative single-particle
energies) are filled up with $N$ particles, leaving $\Omega/2$ upper
levels (with positive single-particle energies) empty~\footnote{This
model is also called Richardson's model, picket-fence model, ladder
model, multilevel pairing model, etc. in the literature.}. It is
worth mentioning that the extension of the exact solution to $T\neq$
0 is not possible at a large value of $\Omega=N$. For the present
schematic model, the number of eigenstates $n_S$, each of which is
$2^{S}$-degenerated, increases almost exponentially with $\Omega$
\begin{equation}
n_S(\Omega)={\rm C}_{N_{\rm pair}}^{\Omega}+\sum_S{{\rm
C}_S^{\Omega}\times{\rm C}_{N_{\rm pair}-\frac{S}{2}}^{\Omega-S}}~,
\label{nos}
\end{equation}
where ${\rm C}_{n}^{m}={m!}/[n!(m-n)!]$ and N$_{\rm pair}=N/2$
is the numbers of pairs distributed over $\Omega$ single-particle
levels. The sum in Eq. \eqref{nos} runs over all the values of total
seniorities S = 0, 2, $\ldots ,\Omega$ . Therefore, at $\Omega=N=$
16 there are 5196627 states, which corresponds to
the order $\sim$ 2.7$\times 10^{13}$ for the square matrix to be
diagonalized. This makes the finite-temperature extension of the
exact pairing solution practically impossible for $\Omega>$ 16 since
all the eigenvalues must be included in the partition function.
Therefore, in the present paper, we limit the calculations up to
$\Omega=N=$ 14, for which there are 73789 eigenstates. For the  GCE
average with respect to the system with $N$ particles and $\Omega=N$
levels, the sum over particle numbers $n$ runs from $n_{min}=1$ to
$n_{max}=2\Omega-1$ with the blocking effect caused by the odd
particle properly taken into account. The calculations are carried
out by using the level distance $\epsilon=$ 1 MeV and the pairing
interaction parameter $G=$ 0.9 MeV. With these parameters, the
values of the pairing gap obtained at $T=$ 0 are equal to around 3,
3.5, and 4.5 MeV for $N=$ 8, 10, and 12 in qualitative agreement
with the empirical systematic for realistic nuclei~\cite{Bohr}.
\subsection{Results within GCE, CE and BCS-based approaches}
\begin{figure}
    \begin{center}
    \includegraphics[width=14cm]{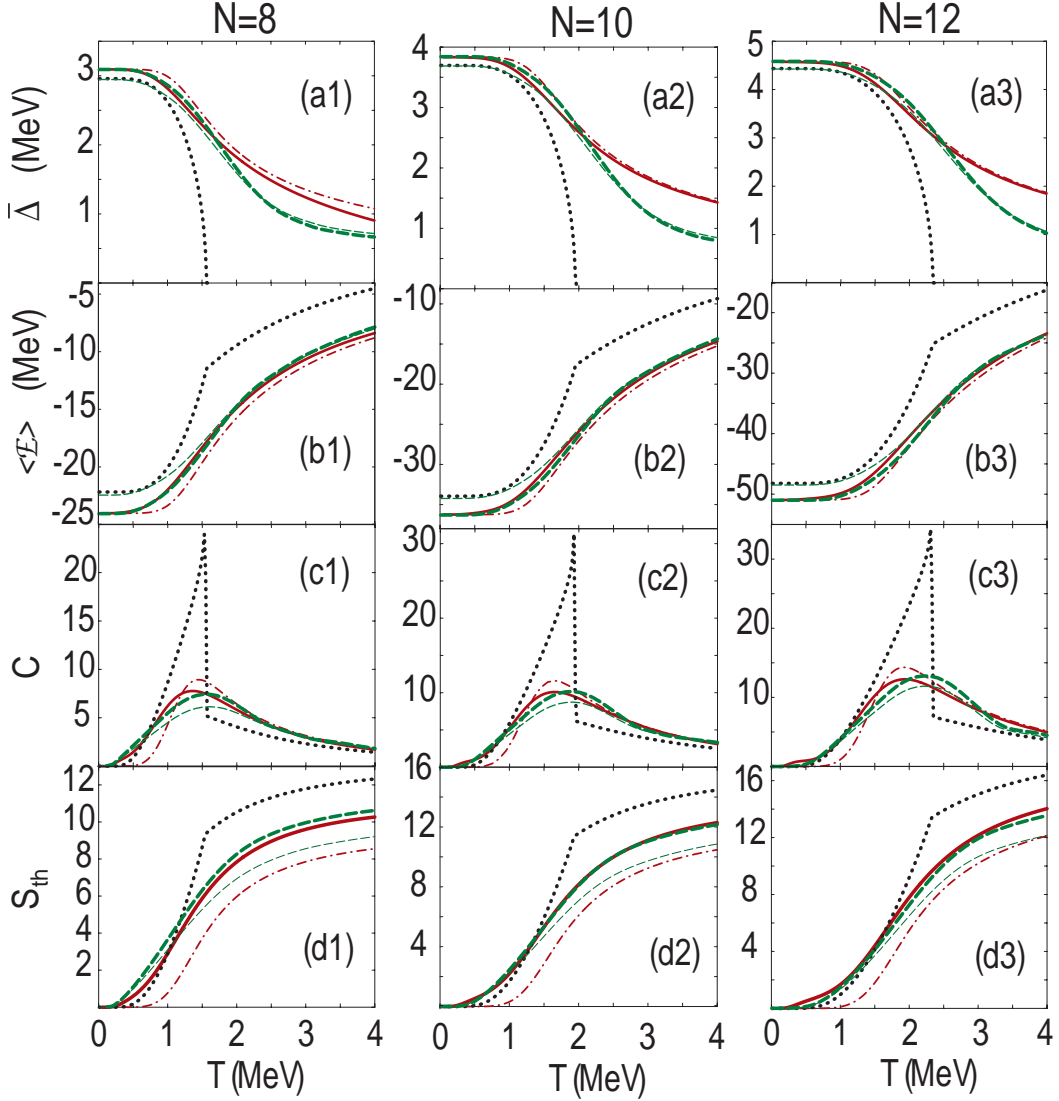}
    \end{center}
 \caption{(Color online) Pairing gaps $\overline\Delta$,
 total energies $\langle{\cal E}\rangle$, heat
 capacities $C$ and thermodynamic entropies, obtained for $N=$ 8, 10, and 12 ($G=$ 0.9 MeV)
 within the FTBCS (dotted lines), FTBCS1+SCQRPA (thin dashed lines),
 FTLN1+SCQRPA (thick dashed lines), CE (dash-dotted lines),
 and GCE (solid lines) vs temperature $T$.\label{D&E&C}}
    \end{figure}
Shown in Fig. \ref{D&E&C} are the pairing gaps, total energies, and
heat capacity obtained for $N=$ 8, 10, and 12 as functions of
temperature $T$ within the FTBCS, FTBCS1+SCQRPA, and FTLN1+SCQRPA,
along with the corresponding results obtained by embedding the exact
solutions (eigenvalues) of the Hamiltonian (\ref{H}) in the CE and
GCE. These latter results using the exact pairing solutions are
referred to as CE and GCE results hereafter. For the FTBCS1+SCQRPA
and FTLN1+SCQRPA the level-weighted gaps $\overline{\Delta}$
(\ref{gapFTBCS1}) are plotted in Figs. \ref{D&E&C} (a1) --
\ref{D&E&C} (a3).

It is seen from this figure that the GCE results are close to the CE
ones for the gaps, obtained from Eq. (\ref{gapGC&C}), as well as for
the total energies, obtained by using Eq. (\ref{EC}). Among
three systems under consideration, the largest discrepancies between
the CE and GCE results are seen in the lightest system ($N=$ 8), for
which the GCE gap is slightly lower than the CE one, and
consequently, the GCE total energy is slightly higher than that
obtained within the CE. As $N$ increases, the high-$T$ values of the
GCE and CE gaps become closer, so do the corresponding total
energies. Different from the BCS results (dotted lines), which show
a collapse of the gap and a spike in the temperature dependence of
the heat capacity at $T=T_{\rm c}$, no singularity occurs in the GCE
and CE results. Both GCE and CE gaps decrease monotonously with
increasing $T$ and remain finite even at $T\gg$ 5 MeV. The
FTBCS1+SCQRPA and FTLN1+SCQRPA predictions for the pairing gap are
found in qualitative agreement with the GCE and CE results [Figs.
\ref{D&E&C} (a1) -- \ref{D&E&C} (a3)]. Because of a different
definition of the pairing gap in the exact solutions within the GCE
and/or CE, where actually no mean-field gap exists (See Sec. II C1), 
one cannot expect a more quantitative agreement between the
predictions by the FTLN1+SCQRPA and the GCE (CE) results. In this
respect, the modified gap (\ref{gapN}) yields a better quantitative
agreement with the GCE (CE) results, as will be seen later in Sec.
\ref{oddevengap}. The two BCS-based approaches
differ noticeably only at $T\leq T_{\rm c}$, where the FTLN1+SCQRPA
gap, due to PNP, practically coincides with the GCE and CE results
at $T <$ 0.5 -- 1 MeV. At $T>T_{\rm c}$ the predictions by two
approaches start to converge to the same value, which decreases with
increasing $T$, remaining smaller than the GCE and CE gaps. 

For the same reason, at $T<T_{\rm c}$, where the total energy predicted by
the FTLN1+SCQRPA agrees very well with the GCE and CE results, the
FTBCS1+SCQRPA energy is significantly larger [Figs \ref{D&E&C} (b1)
-- \ref{D&E&C} (b3)]. At $T>T_{\rm c}$, one finds a remarkable
agreement between the energies predicted by the FTBCS1+SCQRPA,
FTLN1+SCQRPA, and that obtained within the GCE. This seems to be a
natural consequence, given the fact that the two BCS-based
approaches are derived by using the variational procedure within the
GCE. The energies do not suffer either from the difference in the
definitions as the pairing gaps do, as has been discussed above. 
For the heat capacities [Figs \ref{D&E&C} (c1) -- \ref{D&E&C}
(c3)], the spike obtained at $T=T_{\rm c}$ within the FTBCS theory
is completely smeared out within the GCE, CE as well as the
FTBCS1+SCQRPA and FTLN1+SCQRPA, where only a broad bump is seen in a
large temperature region between 0 and 3 MeV. At $T<$ 1 -- 1.2 MeV,
the difference between the GCE and CE energies leads to a
significant discrepancy between the GCE and CE values for the heat
capacity. As the FTBCS1+SCQRPA and FTLN1+SCQRPA heat capacities are
close to the GCE values, this explains the discrepancy reported in
Figs. 4 (b) and 4 (c) of Ref. \cite{FTBCS1} between these results
and the predictions obtained within the CE by the quantum
Monte-Carlo calculations, which use a model Hamiltonian with same
monopole pairing interaction. The thermodynamic entropies $S_{\rm
th}$ are shown as functions of $T$ in Figs. \ref{D&E&C} (d1) --
\ref{D&E&C} (d3). Once again the FTLN1+SCQRPA results for the
thermodynamic entropy $S_{\rm th}$ agree very well with $S_{\rm th}$
obtained within the GCE, whereas such good agreement is seen for the
FTBCS1-SCQRPA results only at $T<$ 1 MeV. The CE thermodynamic
entropy is significantly lower than the values obtained in all other
approaches under consideration, which are based on averaging within
the GCE. On the other hand, the BCS theory strongly overestimates
the thermodynamic entropy at $T>T_{\rm c}$.
\subsection{Results within MCE}
    \begin{figure}
        \includegraphics[width=14cm]{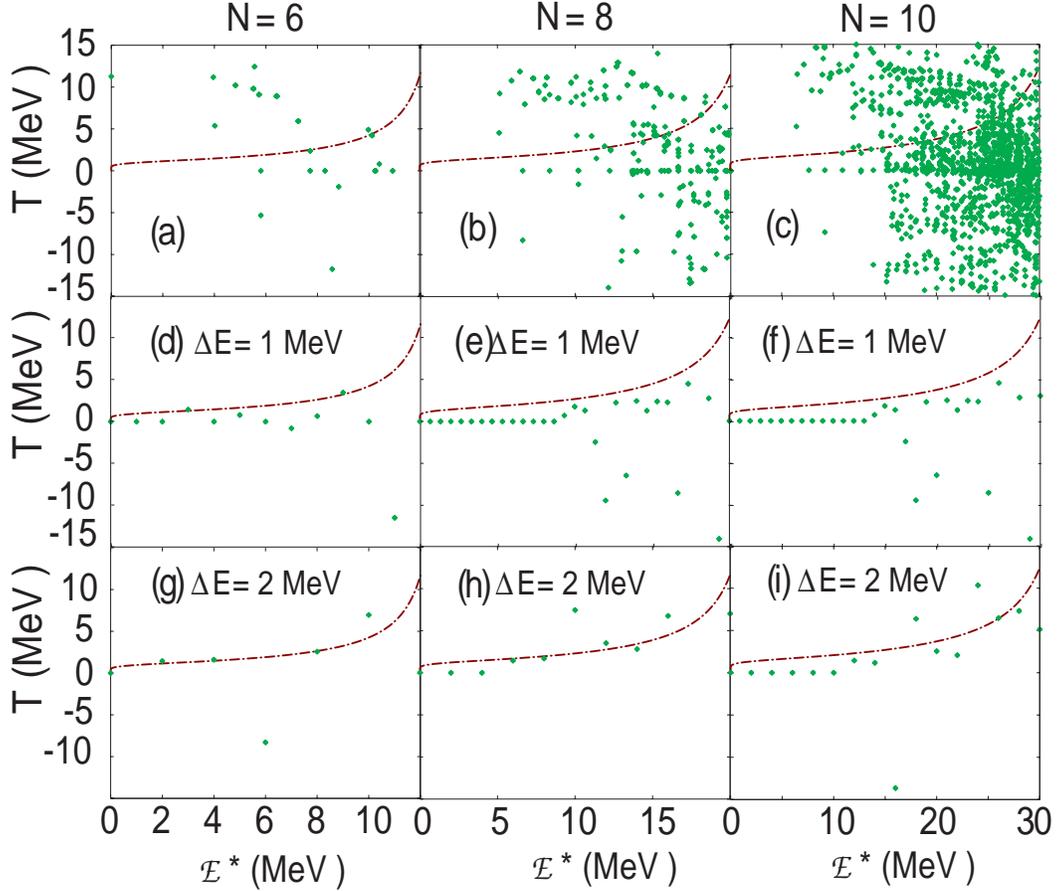}
     \caption{(Color online) Temperature extracted from Eq. (\ref{TMCE})
     within the MCE (dots) vs excitation energy ${\cal E}^{*}$ in
     comparison with the CE results (dash-dotted line) for $N=$ 8, 10,
     and 12. The results in the top panels, (a) -- (c), are
     obtained from the definition of Neumann for entropy Eq. (\ref{Neumann}), whereas those in
     the middle panels, (d) -- (f), and bottom ones, (g) -- (i), are
     calculated from Boltzmann's entropy Eq. (\ref{SME}) with
     two different values of the energy interval
     $\Delta E$ in the statistical weight $\Omega$.
     \label{Tmicro}}
        \end{figure}
The values of temperature within the MCE as extracted by using Eq.
(\ref{TMCE}) are plotted in Fig. \ref{Tmicro} along with the CE
results against the excitation energy ${\cal E}^{*}$. The CE
definition of the latter is ${\cal E}^{*}_{\rm C}\equiv\langle{\cal
E}(T)\rangle_{\rm C}- \langle{\cal E}(T=0)\rangle_{\rm C}$. For
comparison, the results obtained by using the entropy
(\ref{Ss}) are also presented in the top panels [Figs.
\ref{Tmicro} (a) -- \ref{Tmicro} (c)]. They show the values of the
eigenstate temperatures $T_{s}$, which scatter widely around the heat
bath temperature (i.e. the CE result) [Figs. \ref{Tmicro} (a) --
\ref{Tmicro} (c)]. Many of these values are even negative.
Since the eigenstate temperatures $T_{s}$ 
are related with the spread of the exact eigenfunctions
over the ``unperturbed'' basis states $k$ with the weights
$[C_{k}^{(s)}]^{2}$, they need not follow
the trend of the heat bath (or canonical) temperature, which 
depends just on the level density [Eqs. (\ref{SME}) and (\ref{TMCE})].
In fact, with increasing the energy interval $\Delta E$, within which the
levels are counted, the values of the MCE temperature 
determined by Eq.(\ref{TMCE}) gradually converge to the CE
values [Figs. \ref{Tmicro} (a) -- \ref{Tmicro} (i)]. This means
that, the thermal equilibrium within the MCE for the present isolated
pairing model can be reached only at large $N$ and dense spectrum
(small level spacing). Moreover, a full thermalization in 
a system with pure pairing is a subject under question. 
In Ref. \cite{Zele,Volya1}, the thermalization of the system is
characterized by the single-particle temperature, which is obtained by
fitting the occupation numbers of the individual eigenstates to those 
given by the Fermi-Dirac distribution.
The numerical results 
Fig. 12 of Ref. \cite{Volya1}  
show that, the temperature extracted from the
density of states in $^{116}$Sn by using Eq. (\ref{TMCE}) agrees with the
single-particle temperature only when all the residual interactions
are taken into account. With the pure pairing interaction alone, the
single-particle temperature shows a low temperature of the whole
system, whereas the MCE temperature (\ref{TMCE}) is a hyperbola as a
function of the excitation energy with a
singularity in the middle of the spectrum, where it turns negative
[See also Figs. 53 and 54 of Ref. \cite{Zele}].

    \begin{figure}
        \includegraphics[width=14cm]{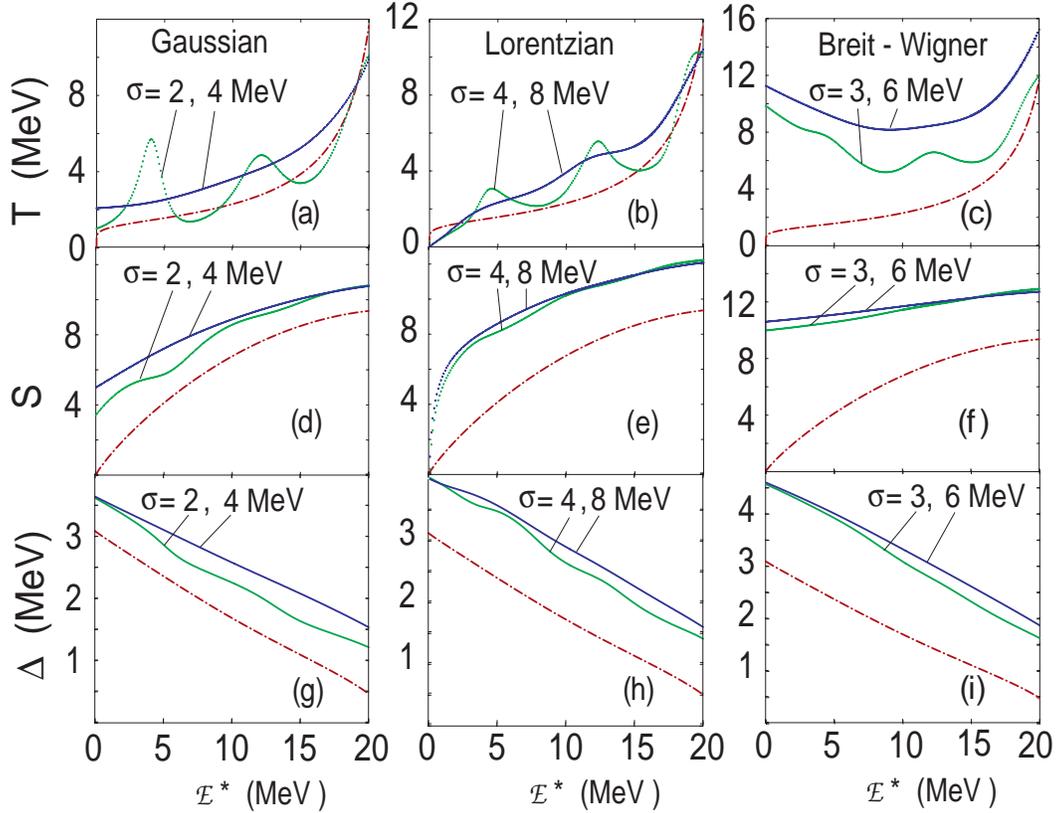}
     \caption{(Color online) Temperatures [(a) -- (c)],
     entropies [(d) -- (f)], and pairing gaps [(g) -- (i)]
     within the MCE as functions of excitation energy ${\cal E}^{*}$
     for $N=$ 8 ($G=$ 0.9 MeV) obtained by using the Gaussian, Lorentz,
     and Breit-Wigner distributions from Eq.
     (\ref{distribution}) for the level density
     at different values of the
     parameter $\sigma$. The dash-dotted lines show the CE values.
     \label{Tfit}}
        \end{figure}
The values of MCE temperature, entropy and gap obtained by using the
Gaussian, Lorentz, and Breit-Wigner distributions from Eq.
(\ref{distribution}) are shown in Fig. \ref{Tfit} as functions of
excitation energy ${\cal E}^{*}$. While the fluctuating behavior of
the microcanonical temperature can be smoothed out by increasing the
parameter $\sigma$ in all three distributions, we found that only
the Gaussian distribution can simultaneously fit both the
temperature and entropy [Figs. \ref{Tfit} (a) and \ref{Tfit} (d)].
The Lorentz distribution can fit only the MCE temperature to the CE
one, but fails to do so for the entropy [Figs. \ref{Tfit} (b) and
\ref{Tfit} (e)], whereas the Breit-Wigner distribution can fit the
MCE temperature to the CE value only at high excitation energies
[Figs. \ref{Tfit} (c)]. A similar result is seen for the pairing
gaps as functions of ${\cal E}^{*}$, where the Gaussian fit gives
the best performance among the three distributions [Compare Figs.
\ref{Tfit} (g), \ref{Tfit} (h), and \ref{Tfit} (i)]\footnote
{The non-vanishing values of $T$ and $S$ at ${\cal E}^{*}=$ 0 in Figs.
\ref{Tfit} (a), \ref{Tfit} (c), \ref{Tfit} (d), \ref{Tfit} (f) are 
the artifacts due to the use of the Gaussian and Breit-Wigner 
distribution functions (14) to smooth out the discrete level density. 
These two distributions have non-zero values $\sim 1/\sigma$ at $x =$ 0, 
where the Lorentz distribution vanishes.}. 
We conclude
that the Gaussian distribution should be chosen as the best one for
smoothing the level density $\rho({\cal E})$ in Eq. (\ref{rho}) to
extract the MCE temperature.
\subsection{Pairing gaps extracted from odd-even mass differences}
\label{oddevengap}
    \begin{figure}
        \includegraphics[width=14cm]{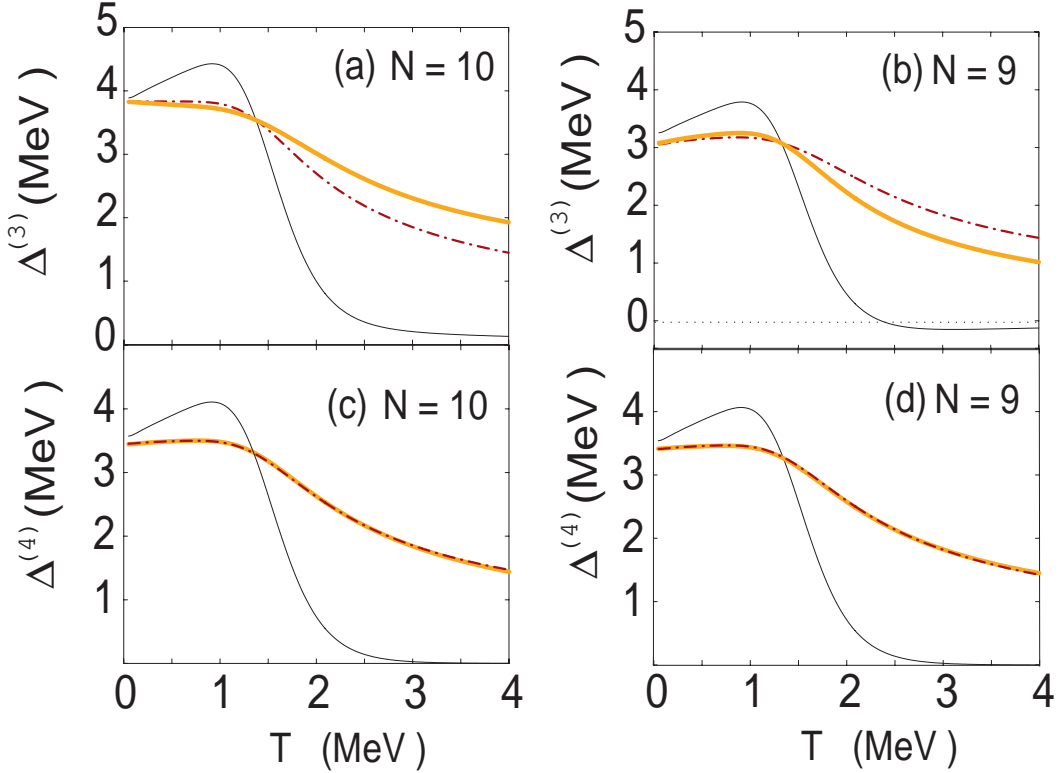}
     \caption{(Color online) Pairing gaps extracted from the odd-even
     mass differences as functions of $T$ for $N=$ 10 (a,c) and $N=$ 9
     (b,d) ($\Omega=$ 10, $G=$ 0.9 MeV). The thin solid and thick solid lines denote the gaps
     $\Delta^{(i)}(\beta, N)$ ($i=$ 3, 4) 
     from Eq. (\ref{Soddeven}), and the modified
     gaps $\widetilde{\Delta}^{(i)}(\beta,N)$ from Eq. (\ref{gapN}), 
     respectively. The dash-dotted lines are the canonical results 
     $\Delta^{(i)}_{\rm C}$.
     The upper panels (a) and (b) show the three-points gaps ($i =$ 3), 
     whereas 
     the the corresponding four-points gaps ($i=$ 4) are shown in the lower
     panels (c) and (d).
     \label{D3}}
        \end{figure}
We extracted the pairing gaps $\Delta^{(i)}(\beta, N)$ ($i=$ 3 and 4) 
by using the simple extension of the odd-even mass formula to $T\neq$ 0 in
Eq. (\ref{Soddeven}) as well as the modified gaps
$\widetilde{\Delta}^{(i)}(\beta,N)$ from Eq. (\ref{gapN}), and
the canonical gaps $\Delta^{(i)}_{\rm C}$ from Eq. (\ref{gapGC&C}) for
several particle numbers up to $N=$ 12. 
In these calculations, the blocking effect caused by the unpaired particle in the
systems with an odd particle number is properly taken
into account in constructing the basis states when diagonalizing the pairing
Hamiltonian.  The results obtained for $N=$ 9 and 10 ($\Omega=$ 10)
are displayed in Fig. \ref{D3}. First of all, the values for $S'$ are found to be always
negative, and increases with $T>$ 1 MeV to reach a 
value of around $-$2 MeV at $T=$ 5 MeV, and vanish at very high $T$. 
By comparing Figs. \ref{D3} (a) and
\ref{D3} (b) one can see a clear manifestation of the parity
effect~\cite{Delft},
which causes the reduction of the three-point gap the in the system
with an odd particle number ($N=$ 9) due to the blocking effect.   
For lights systems as those 
considered here, this reduction is rather strong (about 1 MeV at $T=$ 
0). With increasing $T$, as the thermodynamics weakens the effect of a
single-unpaired particle, the parity effect starts to fade out in such
a way
that the three-point canonical gap $\Delta^{(3)}_{\rm C}(N=9)$
slightly increases with $T$ up to $T\simeq$ 1 MeV, starting from which
it is even slightly larger than $\Delta^{(4)}_{\rm C}(N=9)$. This
feature is found in qualitative agreement with the results obtained in 
Fig. 25 of Ref.
\cite{Delft} for ultrasmall metallic grains.
It is also seen in Fig. \ref{D3} that the
naive extension of the odd-even mass formula to $T\neq$
0, resulting in the gap $\Delta^{(i)}(\beta, N)$ (thin solid lines), fails
to match the temperature-dependence of the canonical gap
$\Delta^{(i)}_{\rm C}$ (dash-dotted lines). The former even increases with
$T$ at $T<$ 1 MeV, whereas it drastically drops at $T>$ 1 -- 1.5
MeV, resulting in a very depleted tail at $T>$ 2 MeV as compared to
the canonical gap $\Delta^{(i)}_{\rm C}$. Moreover, the
three-point gap $\Delta^{(3)}(\beta,N=9)$ even turns negative at $T>$
2.4 MeV, suggesting that such simple extension of the odd-even mass
difference to finite $T$ is invalid. At the same time, the modified
gap $\widetilde{\Delta}^{(i)}(\beta,N)$ (thick solid line) given by Eq.
(\ref{gapN}) is found in much better agreement with the canonical
one. At $T<$ 1.5 MeV, the three point-gap $\widetilde{\Delta}^{(3)}(\beta,N)$
is almost the same as the canonical gap. At
higher $T$, it becomes larger (smaller) than the canonical value for the system 
with an even (odd) $N$, however the
systematic of the
results of our calculations up to $N=$ 12 show we
that this discrepancy decreases with increasing the particle number. The
source of the discrepancy resides in the assumption of the odd-even
mass formula that the gap obtained as the energy difference between
the systems with $N+1$ and $N$ particles is the same as that
obtained from the energy difference between systems with $N$ and
$N-1$ particles. This assumption does not hold for small $N$ (Cf.
Ref. \cite{cons}). The average in the four-point gap 
nearly eliminates 
this difference. As a result, the modified four-point gaps 
$\widetilde{\Delta}^{(4)}(\beta,N)$ practically coincide with the
canonical gaps. A natural consequence of the average in the definition of the 
four-point gap (\ref{4point}) is that the 
gaps obtained in the systems with $N$ and $N-1$ particles are now
nearly the same.  The pairing gaps predicted
by a number of alternative theories~\cite{Moretto,Goodman,Egido,SPA,MBCS},
including those discussed in the present work, are in closer
agreement to the GCE and CE gaps rather than to the gap
$\Delta(\beta,N)$ from Eq. (\ref{Soddeven}). Therefore, the
comparison in Fig. \ref{D3} suggests that formula (\ref{gapN}) is a
much better candidate for the experimental gap at $T\neq$ 0, rather
than the simple odd-even mass difference (\ref{Soddeven}).
\section{Conclusions}
In the present work, a systematic comparison is conducted for
pairing properties of finite systems at finite temperature as
predicted by the exact solutions of the pairing problem embedded in
three principal statistical ensembles, as well as by the recently
developed FTBCS1 (FTLN1)+SCQRPA. The analysis of numerical results
obtained within the doubly-folded equidistant multilevel model for
the pairing gap, total energy, heat capacity, entropies, and MCE
temperature allows us to draw following conclusions.

1) The sharp SN phase transition is indeed smoothed out in exact
calculations within all three principal ensembles. The results
obtained within GCE and CE are very close to each other even for
systems with small number of particles. As for the MCE, although it
can also be used to study the pairing properties of isolated systems
at high-excitation energies, there is a certain ambiguity in the
temperature extracted from the level density due to the discreteness
of a small-size system. This ambiguity, therefore, depends on the
shape and parameter of the distribution employed to smooth the
discrete level density. We found that, in this respect, the normal
(Gaussian) distribution gives the best fit for both of the
temperature and entropy to the canonical values. The wide
fluctuations of MCE temperature obtained here also indicate that
thermal equilibrium within thermally isolated pure-pairing systems
might not be reached. On the other hand, it opens an interesting
perspective of studying the behavior of phase transitions in finite
systems within microcanonical thermodynamics~\cite{Gross} by using
the exact solutions of pairing problem.

2) The predictions by the FTBCS1+SCQRPA and FTLN1+SCQRPA are found
in reasonable agreement with the results obtained by using the exact
solutions embedded in the GCE and CE. The best agreement is see
between the FTLN1+SCQRPA and the GCE results. Once again, this is a
robust confirmation that quasiparticle-number fluctuation, included
in these approximations, is indeed the microscopic origin of the
strong thermal fluctuations that smooth out the sharp SN phase
transition in finite systems.

3) We suggest a novel formula to extract the pairing gap at finite
temperature from the difference of total energies of even and odd
systems where the contribution of uncorrelated single-particle
motion is subtracted. The new formula predicts a pairing gap in
much better agreement with the canonical gap than the simple
finite-temperature extension of the odd-even mass formula.

\acknowledgments

Discussions with Peter Schuck (Orsay) are gratefully acknowledges.
One of us (N.Q.H.) also thanks S. Frauendorf (Notre Dame), and V.
Zelevinsky (East Lansing) for discussions and hospitality during his
visit at the University of Notre Dame and Cyclotron Institute of the
Michigan State University, where a part of this work was presented.

NQH is a RIKEN Asian Program Associate.
The numerical calculations were carried out using the {\scriptsize FORTRAN} IMSL
Library by Visual Numerics on the RIKEN Super Combined Cluster
(RSCC) system.


\begin{thebibliography}{30}
\bibitem{BCS} 
J. Bardeen, L. Cooper, and Schrieffer, Phys. Rev. \textbf{108},
1175 (1957).
\bibitem{Green}N.N. Bogoliubov and S.V. Tyablikov, Sov. Phys. Dokl.
{\bf 4}, 589 (1959) [Dokl. Akad. Nauk. SSSR {\bf 126}, 53 (1959)];
D.N. Zubarev, Sov. Phys. Usp. {\bf 3}, 320 (1960) [Usp. Fiz. Nauk {\bf
71}, 71 (1960); R. Kubo, M. Toda, and N. Hashitsume, {\it Statistical
Physics II - Nonequilibrium Statistical Mechanics} (Springer,
Berlin-Heidelberg, 1985).
\bibitem{Moretto}
L.G. Moretto, Phys. Lett. B \textbf{40}, 1 (1972).
\bibitem{Goodman}
A.L. Goodman, Phys. Rev. C \textbf{29}, 1887 (1984).
\bibitem{Egido}
J.L. Egido, P. Ring, S. Iwasaki, and H.J. Mang, Phys. Lett. B {\bf
154}, 1 (1985).
\bibitem{SPA}
R. Rossignoli, P. Ring and N.D. Dang, Phys. Lett. B \textbf{297}, 9
(1992); N.D. Dang, P. Ring and R. Rossignoli, Phys. Rev. C
\textbf{47}, 606 (1993).
\bibitem{Zele}V. Zelevinsky, B.A. Brown, N. Frazier, and M. Horoi,
Phys. Rep. {\bf 276}, 85 (1996).
\bibitem{MBCS}
N. Dinh Dang and V. Zelevinsky, Phys. Rev. C {\bf 64}, 064319
(2001); N.D. Dang and A. Arima, Phys. Rev. C {\bf 67}, 014304
(2003); N.D. Dang and A. Arima, Phys. Rev. C {\bf 68}, 014318
(2003); N.D. Dang, Nucl. Phys. A {\bf 784}, 147 (2007).
\bibitem{FTBCS1}
N.D. Dang and N.Q. Hung, Phys. Rev. C {\bf 77}, 064315 (2008).
\bibitem{AFTBCS}
N.Q. Hung and N.D. Dang, Phys. Rev. C {\bf 78}, 064315 (2008).
\bibitem{Richardson}
R.W. Richardson, Phys. Lett. {\bf 3}, 277 (1963); Ibid. {\bf 14},
325 (1965); R.W. Richardson and N. Sherman, Nucl. Phys. {\bf 52}, 221
(1964).
\bibitem{EP}A. Volya, B.A. Brown, and V. Zelevinsky, Phys.
Lett. B 509 (2001) 37.
\bibitem{Kubo}R. Kubo, J. Phys. Soc. Japan {\bf 17}, 975 (1962).
\bibitem{Denton}R. Denton, B. M\"{u}hlschlegel, and D.J. Scalapino,
Phys. Rev. B {\bf 8}, 3589 (1973).
\bibitem{Sumaryada}T. Sumaryada and A. Volya, Phys. Rev. C {\bf 76},
024319 (2007).
\bibitem{Bohr}A. Bohr and B.R. Mottelson, {\it Nuclear structure} {\bf
I} (Bejamin, NY, 1969).
\bibitem{Ring}P. Ring and P. Schuck, {\it The Nuclear Many-Body
Problem} (Springer, Heidelberg, 2004).
\bibitem{Delft}J. von Delft and D.C. Ralph, Phys. Rep. {\bf 345}, 61 (2001).
\bibitem{Ross}R. Rossignoli, N. Canoza, and P. Ring, Ann. Phys. {\bf
275}, 1 (1999).
\bibitem{Frau}S. Frauendorf, N.K. Kuzmenko, V.M. Mikhajlov, and J.A.
Sheikh, Phys. Rev. B {\bf 68}, 024518 (2003).
\bibitem{BW}G. Breit, Phys. Rev. {\bf 58}, 506 (1940);
E.P. Wigner, Phys. Rev. {\bf 70}, 15 (1946); M. Danos and W. Greiner,
Phys. Rev. {\bf 134}, B284 (1964).
\bibitem{Kaneko}K. Kaneko and M. Hasegawa, Phys. Rev. C {\bf 72}, 024307 
(2005), K. Kaneko {\it et al.}, Phys. Rev. C {\bf 74}, 024325 (2006).
\bibitem{Bogo}N.N. Bogoliubov, JETP {\bf 34}, 58 (1958).
\bibitem{SCQRPA}N.Q. Hung and N.D. Dang, Phys. Rev. C {\bf 76}, 054302
(2007); {\bf 77}, 029905(E) (2008).
\bibitem{LN}H.J. Lipkin, Ann. Phys. {\bf 9}, 272 (1960); Y. Nogami,
Phys. Rev. {\bf 134}, B313 (1964);
H.C. Pradhan, Y. Nogami, and J. Law, Nucl. Phys. A {\bf 201}, 357
(1973).
\bibitem{PNP}H. Olofsson, R. Bengtsson, P. M\"{o}ller, Nucl. Phys. A
{\bf 784}, 104 (2007).
\bibitem{PMBCS}N. Dinh Dang, Phys. Rev. C {\bf 76}, 064320 (2007).
\bibitem{Volya1}A. Volya, V. Zelevinsky, and B. Alex Brown, Phys. Rev.
C {\bf 65}, 054312 (2002).
\bibitem{cons}N. Dinh Dang, Phys. Rev. C {\bf 74}, 024318 (2006).
\bibitem{Gross}D.H.E. Gross and J.F. Kenney, J. Chem. Phys. {\bf 112},
224111 (2005).
\end{thebibliography}
\end{document}